# Design of Prototype Front-end Digital Module for Direct Dark Matter Detection based on LAr

X. Zhu, Z. Shen, C. Feng, S. Liu, Q. An

*Abstract*— Liquid argon (LAr) with a high light yield of approximately 40 photons per keV is an attractive target for the direct detection of weakly interacting massive particles (WIMPs), which are well motivated galactic dark matter candidates. We studies a front-end electronics design for a LAr dark matter detector with the scintillation read out by PMT, which has an input dynamic range from 5pC to 1nC, and high resolution that single photoelectron can be distinguished. In this paper, we present the design of front-end digital module (FDM) which is the important portion of electronics. The prototype FDM is equipped with 14-bit 1-GSPS analog-to-digital converters (ADC), and the performance of prototype FDM had been test in lab (e.g. ENOB is 10.40 bits @298 MHz). Moreover, this prototype FDM had been tested with LAr detector collaboratively and test results are also presented.

## I. INTRODUCTION

WEAKLY interacting massive particle (WIMP) is a well-motivated galactic dark matter candidate. Numerous direct detection experiments are being developed to detect WIMPs [1]-[3]. Liquid argon (LAr) detectors, with a high light yield of approximately 40 photons per keV, are attractive detectors for the direct detection of WIMPs [4].

Liquid argon provides outstanding pulse-shape discrimination (PSD) based on scintillation timing. The excitation and ionization of the medium from particles interacting with argon atom will lead to two excited state. The two excited states have different lifetimes, about 6 ns for the singlet state and 1.6 μs for the triplet state [5]-[6]. Utilizing PSD method, we can discriminate the nuclear recoil events from electron-induced background events [7]-[8]. In order to obtain good discriminate ability, we are supposed to design the front-end electronics with high speed and high accuracy. The front-end electronics are designed to simultaneously read out approximately 60 photomultiplier tubes (PMTs) which combined with about 1-ton liquid argon. And these front-end electronics have an input dynamic range from 5pC to 1nC, while also have high resolution that single photoelectron can be distinguished.

The front-end electronics are consist of preamplifier card, front-end digital module (FDM), and trigger clock module (TCM). In this paper, we investigated the prototype FDM to satisfy the requirement of LAr detector.

The remainder of this manuscript is as follows: The design of prototype FDM is presented in section II. Section III introduces the test of the FDM performance. The result of test with LAr detector is shown in section IV. Section V concludes the manuscript.

## II. FRONT-END DIGITAL MODULE

### A. Basic structure

Signals from the PMTs are processed by preamplifier cards then sent to FDM. Two 14-bit 1-GSPS analog-to-digital converters (ADCs) are integrated on FDM. The Field Programmable Gate Array (FPGA) acquires the data from ADCs and transmits the information to TCM by differential star pairs on PXIe chassis. This information contains number of hits and sum of energy pre-processed by FPGA. TCM processes the information from each FDM to obtain trigger information. Moreover, Whether FDM packages the data from ADC depends on the trigger information from TCM. Fig.1 shows the basic structure of the Front End Electronics.

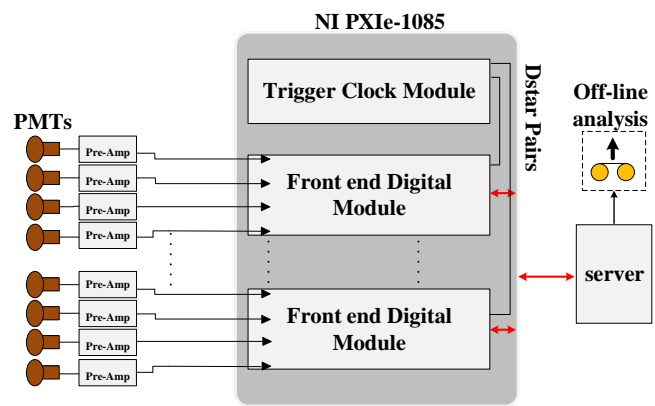

Fig. 1. Basic structure of front end electronics.

Before the pre research work begins, we were supposed to do some feasibility verification. Therefore, we designed a prototype FDM, and tested the performance with the detector.

### B. Design of prototype FDM

The analog signal is received with SMA connector and converted from single-ended to differential by a transformer (TC1-1-13MA+), and then digitized by a 14-bit 1-GSPS ADC (AD9680). The ADC outputs were sent to an FPGA chip

This work was supported by the National Key Technologies R&D Program (Grant No. 2016YFA0400300), and the fundamental Research Funds for the Central Universities (Grant No. WK2030040097).

Xing Zhu, Zhongtao Shen, Changqing Feng, Shubin Liu, and Qi An are with State Key Laboratory of Particle Detection and Electronics, University of Science and Technology of China, No.96, Jinzhai Road, Hefei, Anhui, China (e-mail: henzt@ustc.edu.cn). And the corresponding author is Zhongtao Shen.

(XC7K410T) via high speed link (JESD204B). And the original data of signal waveform is transmitted to computer by a USB2.0 port. The block diagram is shown in fig.2.

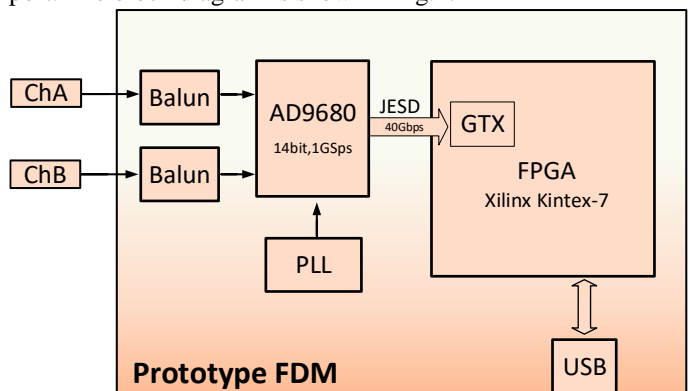

Fig. 2. Block diagram of prototype FDM

### III. PERFORMANCE OF PROTOTYPE FDM

In order to acquire the LAr detector signals with enough details, the FDM should accurately sample the rising edge of waveform which is approximately 5 ns.

#### A. Performance Test platform

Effective number of bits (ENOB) of the prototypes was tested in the laboratory. The electronics performance test setup was shown in the Fig.3. An Agilent E4438C Vector Signal Generator, with a series of narrow-band filters, were used as a sine wave signal source, to perform the evaluation test of the FDM. We acquired the data through USB2.0, then calculated the ENOB.

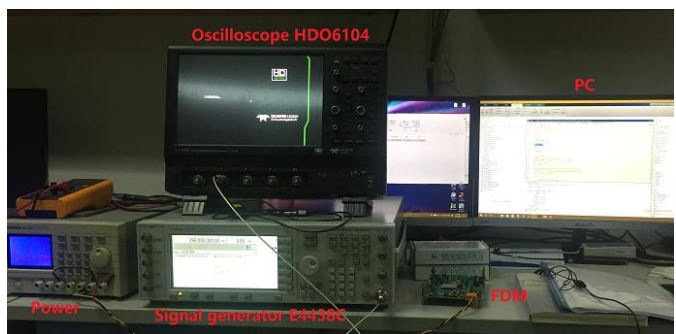

Fig. 3. Electronics performance test setup

#### B. Test results of performance

Fig. 4 is the spectrum of sine wave while the frequency of input signal is 298 MHz. The noise and harmonics are in a quite low level from the spectrum. Table. I shows the ENOB of the sampling board, which is better than 9.91 bits for up to 488 MHz high speed signals, which is consistent with the AD9680 datasheet. And the range of PMTs signal is from 0 Hz to 200MHz. This tests show that our prototype FDM can satisfy requirement of LAr detector.

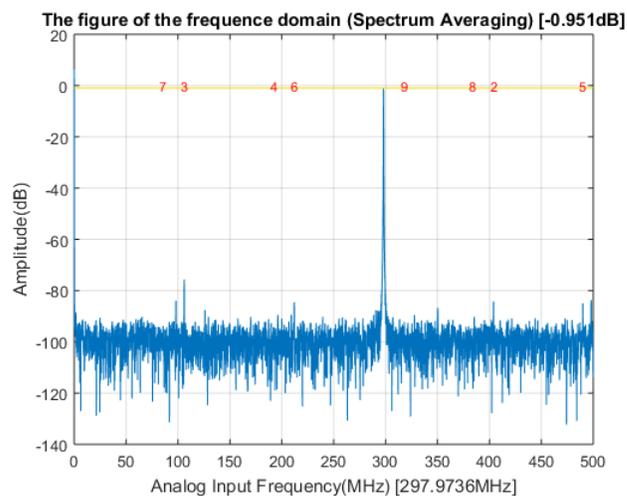

Fig. 4. Spectrum of sine wave. The input frequency is 298 MHz. And Enob is 10.40 bits.

**Table I:** ENOB of FDM at 14-bit 1-GSps

| $F_{in}$/MHz | 30.5 | 98 | 175 | 298 | 378 | 488 |
|---|---|---|---|---|---|---|
| ENOB/bit | 10.35 | 10.57 | 10.18 | 10.40 | 9.97 | 9.91 |

### IV. TEST WITH PMT

The prototype FDM was tested with PMT to assess the ability in the Institute of High Energy Physics of the Chinese Academy of Sciences, as shown in Fig. 5.

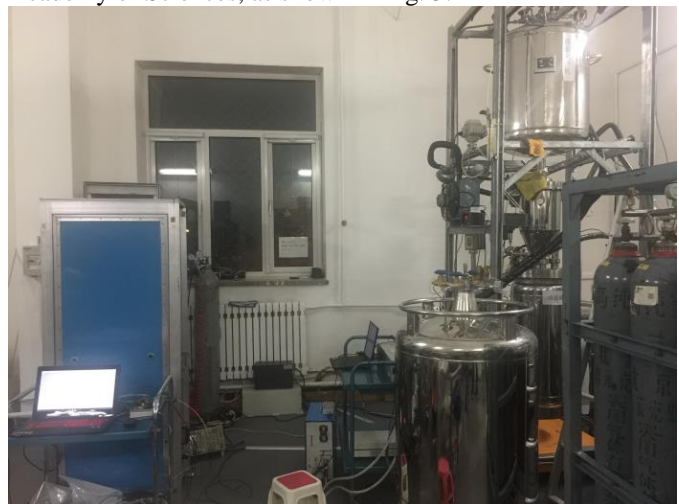

Fig. 5. Experimental setup in IHEP.

#### A. Single-photoelectron experiment

To verify the resolution of FDM we collected waveform of single photoelectron. Fig. 6 and Fig. 7 respectively shows waveform and spectrum. As shown in this figures, the prototype FDM has enough accuracy to distinguish single photon signal.

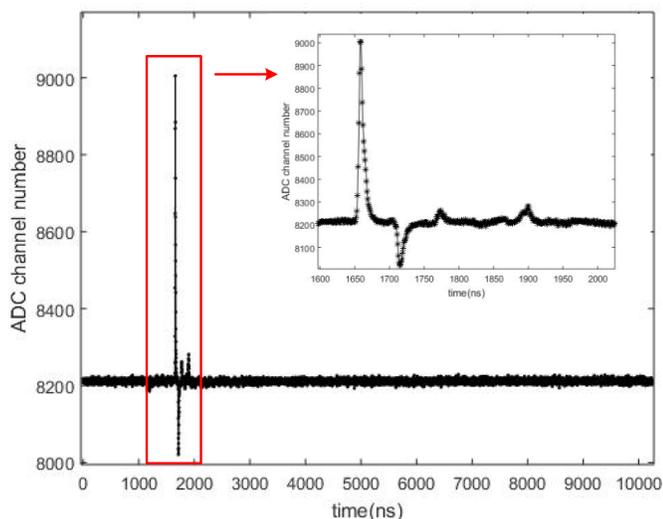

Fig. 6. The waveform of single photoelectron.

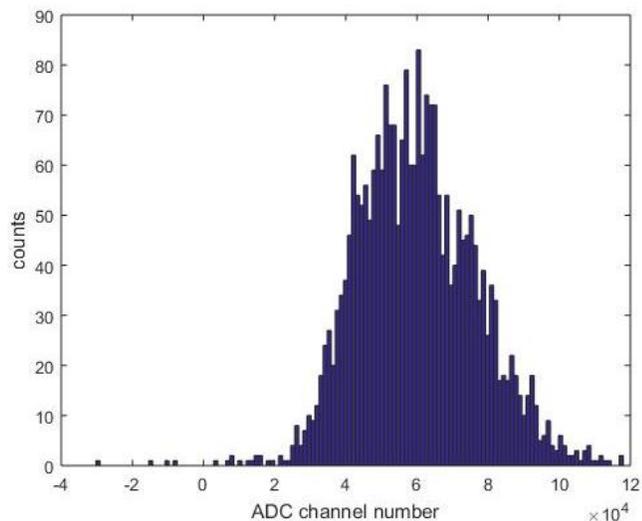

Fig. 7. The spectrum of single photoelectron.

### B. Waveform of LAr detector

We also did some tests of LAr detector. The LAr detector was exposed to a $^{22}$Na gamma source. The waveform of LAr detector acquired by FDM is shown in Fig. 8.

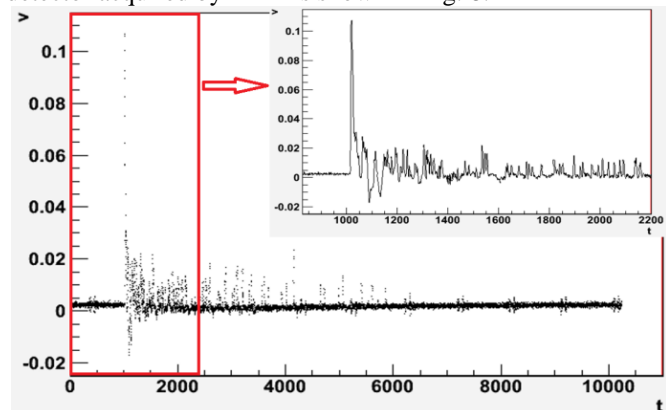

Fig. 8. The waveform of LAr detector

## V. CONCLUSION

A prototype FDM are designed, and detailed tests of FDM have been done. According to the test results, the module has been proven to perform well. This Front End Electronics can bring better performance in PSD to detect rare nuclear recoil events.

### ACKNOWLEDGEMENT

The authors are thankful to Dr. M. Guan at the Institute of High Energy Physics of the Chinese Academy of Sciences for providing experimental facilities and help.